\documentclass[iop]{emulateapj}

\newcommand{\msun}{M$_{\odot}$}
\newcommand{\um}{$\mu$m}

\newcommand{\sig}{$\sigma$}

\newcommand{\clfind}{{\tt clfind2d}}%

\begin{document}

\title{Measuring the Clump Mass Function in the Age of SCUBA2, Herschel, 
and ALMA}

\author{Michael A. Reid}
\affil{Department of Astronomy and Astrophysics, University of Toronto, Toronto, ON, M5S 3H4, Canada}
\author{James Wadsley, Nicolas Petitclerc, \& Alison Sills}
\affil{Department of Physics and Astronomy, McMaster University, 1280 Main St. W., Hamilton, ON, L8S 4M1, Canada}

\begin{abstract}

	We use simulated images of star-forming regions to explore the 
effects of various image acquisition techniques on the derived clump 
mass function.  In particular, we focus on the effects of finite image 
angular resolution, the presence of noise, and spatial filtering.  We 
find that, even when the image has been so heavily degraded with added 
noise and lowered angular resolution that the clumps it contains clearly 
no longer correspond to pre-stellar cores, still the clump mass function 
is typically consistent with the stellar initial mass function within 
their mutual uncertainties.  We explain this result by suggesting that 
noise, source blending, and spatial filtering all randomly perturb the 
clump masses, biasing the mass function toward a lognormal form whose 
high-mass end mimics a Salpeter power law.  We argue that this is a 
consequence of the central limit theorem and that it strongly limits our 
ability to accurately measure the true mass function of the clumps.  We 
support this conclusion by showing that the characteristic mass scale of 
the clump mass function, represented by the ``break mass'', scales as a 
simple function of the angular resolution of the image from which the 
clump mass function is derived.  This strongly constrains our ability to 
use the clump mass function to derive a star formation efficiency.  We 
discuss the potential and limitations of the current and next generation 
of instruments for measuring the clump mass function.

\end{abstract}

\keywords{ISM: structure --- methods: data analysis --- stars: formation 
--- stars: mass function --- submillmeter}

\section{Introduction}
	
	The mass function of molecular cloud clumps is increasingly 
being used as a tool to test theories of star formation.  By comparing 
the mass function of pre-stellar molecular cloud cores to the initial 
mass function (IMF) of stars, one hopes to constrain things 
like the efficiency and timescales of star formation.  Similarly, one 
can test different theories of star formation by comparing the clump 
mass functions they predict to the observations.  The potential for 
using the clump mass function as a diagnostic of massive star formation 
is particularly attractive.  There is no clear one-to-one relationship 
between individual massive stars and individual pre-stellar molecular 
cloud cores.  Massive stars may form by competitive accretion, by 
monolithic collapse of a molecular cloud core, or some combination.  
Presumably these two theories could be distinguished on the basis of the 
clump mass functions they predict.  

	In recent years, measurements of the shape of the clump mass 
function in nearby low-mass star-forming regions have demonstrated good 
agreement with the shape of the stellar initial mass function 
\citep{ts98,man98,m01,dj2000,dj2001,tot02}.  The slopes of the high-mass 
ends of the stellar IMF and the clump mass function agree within their 
uncertainties.  The two mass functions peak at different masses, but 
this is taken to be indicative of the star formation efficiency.  These 
quantitative resemblances between the stellar IMF and the clump mass 
function have been interpreted as evidence that the clumps we observe 
are the direct precursors of individual low-mass stars (or low-order 
multiples).

	Our ability to use the clump mass function as a test of theories 
of star formation hinges crucially on our ability to measure it 
accurately and interpret it confidently.  In this paper, we will argue 
that the observations to date have not provided definitive measurements
of either the shape or the characteristic mass scales of the clump mass 
function.  We will also argue that this situation may be about to 
change, thanks to upcoming observations to be made with instruments such 
as the Submillimetre Common-User Bolometer Array 2 (SCUBA2, 
\citealt{scuba2}) on the James Clerk Maxwell Telescope (JCMT) as well as 
the Spectral and Photographic Imaging Receiver (SPIRE, \citealt{spire}) 
and the Photodetector Array Camera and Spectrometer (PACS, 
\citealt{pacs}) on the Herschel Space Observatory.

	\citet{rw06b} began this investigation by showing that the 
interpretation of observational clump mass functions is biased by the 
effects of small-number statistics and certain fitting techniques.  
\citet{rw06b} hypothesized that the observed shape of the clump mass 
function may be determined as much by our observational and analytical 
techniques as by the physics of star-forming clouds.  They showed that a 
lognormal shape provides the best fit, with the fewest parameters, to 
the observed clump mass functions, but they could not draw hard 
conclusions about the origin of this lognormal shape.  In this paper, we 
will extend this analysis, arguing that our observational techniques 
play a role in determining the functional form of the clump mass 
function, biasing it toward a lognormal shape. 

\section{Analysis}

\subsection{Problems in the Interpretation of Clump Mass Functions}
\label{sec:define}

	From an observational perspective, there is a very important 
distinction between stars and clumps.  Stars appear as points whose 
luminosities can be measured accurately, save for relatively small, 
quantifiable uncertainties due to binarity and crowding.  Clumps, on the 
other hand, are extended objects viewed against a background of emission 
from their parent molecular clouds and from Galactic cirrus.  The flux 
one measures from a clump can depend strongly on, among other things, 
the noise level in the image, the angular resolution of the image, 
projection effects, image processing techniques that might be used 
during or after data acquisition, and the choice of clump-finding 
algorithm.  Each of these effects varies in magnitude with the mass and 
internal structure of the clump so the whole mass function is not evenly 
affected.

	To interpret a clump mass function, it is necessary to know how 
it was measured.  The most common method is to measure the clump mass 
function from images of thermal dust emission at millimeter and 
submillimeter wavelengths.  Another common method uses molecular line 
maps analysed using routines such as GAUSSCLUMPS 
(e.g.~\citealt{kramer98}, \citealt{sg90}) or, more recently, dendrogram 
methods \citep{ros08}.  A further method extracts the clumps from dust 
extinction maps \citep{alves07}.  We concentrate on the dust 
continuum observations here, both because of their popularity in 
the past and because they will be produced in quantity by 
instruments such as Herschel, SCUBA2, and ALMA.

	Maps of the dust continuum emission in star-forming regions show 
a combination of discrete clumps and smooth, often filamentary 
structures.  These structures are sectioned into clumps either by eye 
\citep{ts98,tot02} or using one of a variety of algorithms, such as 
\clfind\ \citep{clfindref}.  Recently, \clfind\ has been criticized for 
its inability to, among other things, produce accurate mass functions in 
regions where the emission is crowded or has a lot of structure on 
multiple spatial scales \citep{p09,k10,cr10}.  These are reasonable 
criticisms of \clfind\ but, as we will argue throughout the rest of this 
section, our concerns about the interpretation of the clump mass 
function apply to many different clump-finding algorithms.

	Clump mass functions are subject to several uncertainties which 
cannot be addressed easily with existing observations. Probably the most 
significant among these is the conversion of flux to mass.  This 
conversion depends on several parameters, including the emissivity, 
temperature, and opacity of the dust, which are not well characterized 
and probably vary both from clump to clump and within clumps.  
Typically, estimated values are assumed to hold for all clumps in the 
sample.  However, it will be important to remember that the assumption of 
standardized values for the dust emissivity, temperature, and opacity 
and their application to the entire volume of each clump constitutes an 
essentially random perturbation to the mass of each clump.  These 
``random'' errors may be systematic in the sense that they bias all of 
the clump masses in the same direction--either too high or too low--but 
they will affect each clump's mass by a different, unknown amount.  The 
cumulative effect on the clump mass function of many such random errors 
may be significant.

	The choice of clump-finding algorithm constitutes an additional 
uncertainty which is difficult to control.  Any given algorithm may 
assign too much or too little of the background emission in the image to 
a given clump.  Given that clump extraction algorithms typically do not 
pair submillimeter continuum maps with velocity information, many of 
these algorithms will identify gravitationally unbound clumps.  Some 
attempts to exclude gravitationally unbound clumps is usually made, but 
these attempts depend on some knowledge of the internal structures of 
the clumps, which is also lacking.

	Uncertainties due to the inclusion of clumps which have already 
formed one or more stars (i.e. those which are not ``pre-protostellar'') 
can be mitigated when deep mid- and far-infrared data are available, but 
this has not historically been the case.  

	Superposition of clumps and filaments along the line of sight 
represents a final significant source of uncontrolled uncertainty.  
Superposition will tend to increase the measured masses of clumps but, 
again, by random, unknown amounts.
	
	Little can be said so far about the cumulative quantitative 
effects of these uncertainties.  Upcoming multi-wavelength observations 
with Herschel and SCUBA2 will help address uncertainties due to spatial 
variations in dust properties and afford a better understanding of the 
structures of clumps in nearby star-forming regions.  The next 
generation of clump finding algorithms promise to improve the accuracy 
with which clump structures can be determined and their masses 
calculated.  Still, there will continue to exist significant 
unquantified uncertainties.

	In this paper, we set aside all of the aforementioned concerns 
and instead seek to quantify the effects of our image \emph{acquisition} 
techniques on the clump mass function.  Image acquisition techniques 
vary substantially from one telescope to another, but they essentially 
all consist of combinations of a small set of elemental processes.  
Images can have more or less noise, coarser or finer angular resolution, 
and they can be spatially filtered in several ways (e.g. chopping and 
interferometry).  Each of these has some effect on measurements of the 
clump mass function.  In this paper, we seek to explore and quantify 
these effects.
		
\subsection{The Reference Image}
\label{sec:refimg}
	
	Our goal is to evaluate accurately the effects on the clump mass 
function of image noise, limited angular resolution, and spatial 
filtering.  We thus begin with a simulated image which has no 
instrumental noise, higher intrinsic angular resolution than any we wish 
to simulate, and flux on a wider range of spatial scales than those we 
wish to simulate.  We call this image our `reference image'.
	
        Our reference image was derived from a snapshot of a 
star-forming region simulated using smoothed-particle hydrodynamics.  
The simulation began with a spherical cloud of turbulent, isothermal gas 
at 10~K which was allowed to collapse under the influence of its own 
gravity.  This has been a standard approach in the last several years 
(e.g.~\citealt{bate03}). The simulation was large, with a mass similar 
to that inferred for typical star forming clouds (5000 \msun\ with 36M 
particles).  The simulated cloud was marginally bound, having an rms 
Mach number of 13.42 and an initial radius of 4~pc.  As our primary 
interest was pre-stellar cores, sink particles were not used.  The 
resolution (gravitational and hydrodynamical) was limited to 50~AU, 
corresponding to a maximum number density of $\sim 4\times 10^7$ 
cm$^{-3}$ (mean molecular weight 2.33) and a minimum Jeans Mass of $\sim 
0.04$ \msun (330 particles). The simulation was performed in parallel 
using the GASOLINE code \citep{wadsley} and is described in more detail 
in \cite{petitclerc}.  Our reference image was developed from a snapshot 
of the simulation taken 130,000 years after the cloud began collapsing.  
That is the epoch at which dense cores have appeared but before 
radiative effects become significant on the spatial scales of interest 
to us.

\begin{figure}
\begin{center}
\includegraphics[width=\columnwidth]{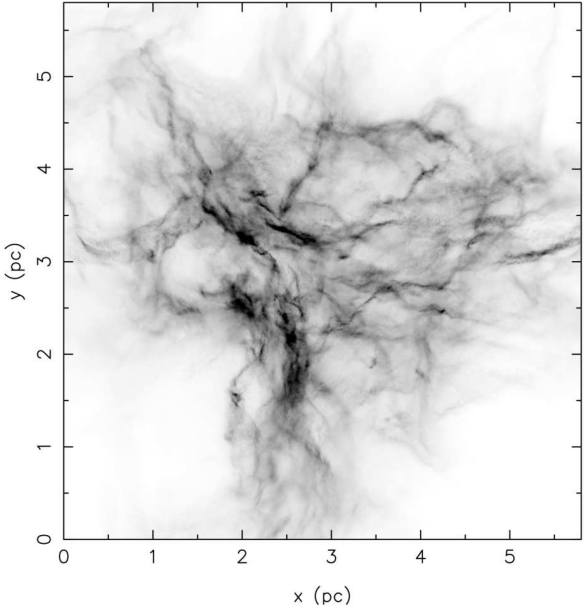}
\caption{Column density map made by projecting our simulated 
the star-forming region along its z axis. Darker areas 
indicate higher column densities. \label{fig:simimg}} 
\end{center} 
\end{figure}

	We are interested in simulating the results of observations 
which produce two-dimensional images of (presumed) optically thin 
emission.  Thus, we made our image by projecting the simulation along a 
randomly chosen axis, neglecting optical depth effects.  This reference 
image is shown in Figure~\ref{fig:simimg}.  All subsequent images used 
in our analysis are modified versions of this one.

\subsection{The Reference Clump Mass Function}
\label{sec:refcmfsec}
	
	Instruments such as SCUBA2 and Herschel are producing images 
containing so many clumps and sitting on such complex backgrounds of 
Galactic cirrus and extragalactic point sources that manual clump 
identification is now time-prohibitive.  Manual source extraction is 
also undesirable because it does not produce consistent, repeatable 
results and therefore introduces an unnecessary variable in the study of 
the clump mass function.  A huge variety of source extraction tools are 
being developed to handle these new types and volumes of data.  For the 
reasons discussed previously, we will use \clfind\ for our analysis, but 
we have taken care to use it in a consistent, automatic way to eliminate 
the variable of manual ``tuning'' of the results.  Our method consists 
of measuring the rms noise, \sig, in emission-free parts of the image 
and then setting the threshold and contour intervals in \clfind\ to 
3\sig\ and 2\sig\ respectively.  

\begin{figure} 
\begin{center} 
\includegraphics[width=\columnwidth,angle=270]{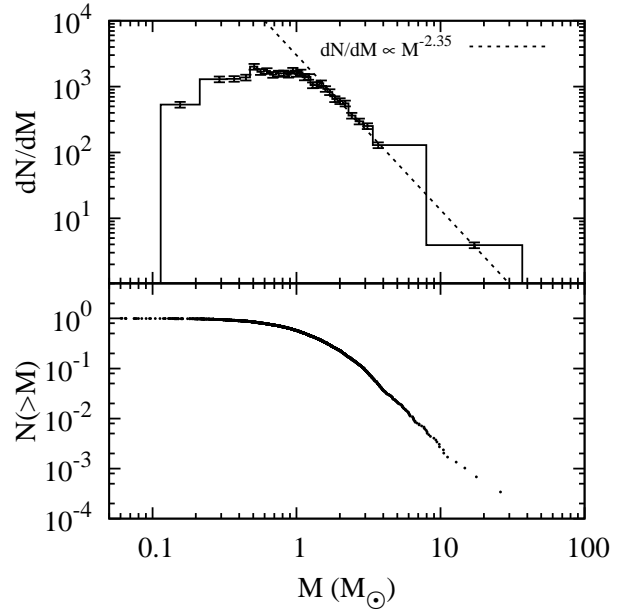} 
\caption{Differential (\emph{upper panel}) and cumulative (\emph{lower 
panel}) mass functions of the cores in the reference image, shown in 
Fig.~\ref{fig:simimg}.  The error bars are too small to be represented 
clearly in the cumulative mass function.  The DCMF
has a constant 100 objects per bin.  The dashed lines in the upper 
panel indicate a mass function with a Salpeter slope, $dN/dM 
\propto M^{-2.35}$.\label{fig:refcmf}} 
\end{center} 
\end{figure}

	We always produce two different representations of the clump 
mass function: the differential clump mass function (DCMF), $dN/dM$, and 
the cumulative clump mass function (CCMF), $N(>M)$.  In producing the 
DCMF for a given image, we use a fixed number of clumps per bin (rather 
than fixed bin widths), following the prescription of \citet{ubeda} for 
producing reliable histograms.  The DCMF and CCMF for the clumps 
extracted from the reference image in this manner are shown in 
Figure~\ref{fig:refcmf}.  We note that the uncertainties in the 
flux-to-mass conversion, discussed in \S\ref{sec:define}, affect the 
observations but not our simulations, which track mass directly.

	The evidence so far suggests that the clump mass function, like 
the stellar initial mass function, follows a Salpeter-like power-law of 
the form $dN/dM \propto M^{-\alpha}$ with $\alpha \simeq 2.35$ above 
about $0.1$--$1$ \msun.  For comparison, we have drawn such a power-law 
on the DCMF in Figure~\ref{fig:refcmf}.  As is evident, the DCMF of the 
simulated clumps in their `native' form is well-fit by a Salpeter-like 
power-law.  Hence, we conclude that our simulations replicate the 
observations well enough to form the basis of our analysis.

\subsection{Resolution Effects}
\label{sec:reseffects}
	
	We first assess the effect on the clump mass function of 
limiting the angular resolution of the observations.  The historically 
relatively coarse angular resolution of submillimeter telescopes has 
produced images in which individual pre-stellar cores may be convolved 
together.  Interferometers achieve high angular resolution at the 
expense of filtering out flux on potentially important spatial scales. 
In this section, we concern ourselves only with the single-dish case of 
degrading the angular resolution while conserving total flux.
	
	We might expect that degrading the angular resolution of an 
image would simply convolve low-mass clumps together, artificially 
inflating the number of high-mass clumps and lowering the number of 
low-mass clumps.  If this were so, we would expect the slope of the mass 
function to become progressively shallower as the resolution was 
degraded.  However, convolution is more complicated than this.  
Convolution mainly blends \emph{small} clumps together; small clumps can 
have a range of masses, not always low.  Also, the effects of 
convolution are much more significant where clumps are crowded together 
and crowding is more common among intermediate- and high-mass clumps 
than among low-mass ones.  Finally, even in a noise-free image, 
convolution can combine peaks which were formerly below the clump 
detection threshold and raise them above it, creating new low-mass 
clumps.

	To test the effects of degraded angular resolution on the clump 
mass function, we produced several versions of the reference image, each 
with progressively coarser angular resolution.  We cast our discussion 
in terms of varying the distance to the simulated region, but we remind 
the reader that this is fully equivalent to varying the size of the 
telescope used to observe it.

\begin{figure*} 
\begin{center} 
\includegraphics[width=1.5\columnwidth,angle=270]{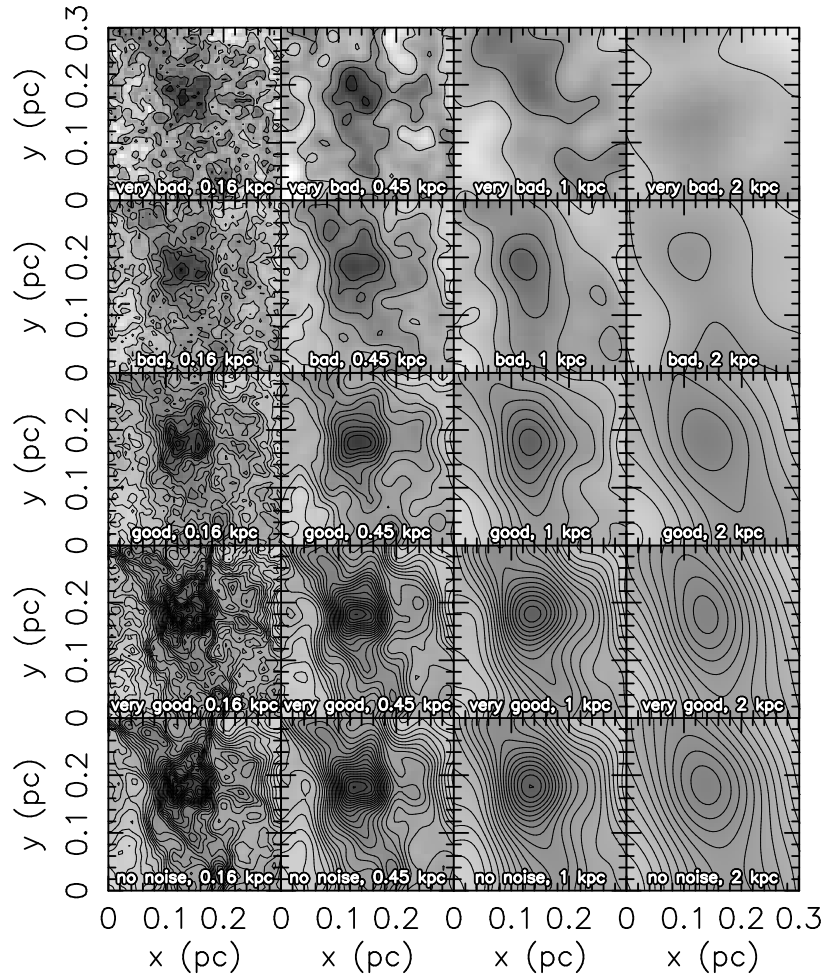} 
\caption{Representations of the reference image as seen at four 
different angular resolutions (\emph{x axis}) and five different noise 
levels (\emph{y axis}).  Each panel shows the same representative 
0.3 pc $\times$ 0.3~pc section of the 
reference image.  In each panel, the contours are spaced at intervals 
of 2\sig\ and they correspond to the contours used by \clfind.  Each 
image is shown as it would appear if observed with a 14\arcsec\ circular 
beam from the distance shown in each panel.  The noise levels are 
described in the text.
\label{fig:resnoise}} 
\end{center} 
\end{figure*}

	To change the resolution of our images, we convolve them with a 
circular Gaussian beam.  We chose convolved resolutions intended to 
mimic those that will be obtainable with SCUBA2 on the JCMT, observing 
at 850~\um, although of course the trends observed in this analysis 
would hold for any single dish telescope mapping dust continuum emission 
in a similar way.  At this wavelength, the beam size of the JCMT is 
about 14\arcsec.  Using this beam size, we simulate observations of 
star-forming regions at distances of 160~pc, 450~pc, 1~kpc, and 2~kpc, 
corresponding respectively to the distances of the $\rho$~Ophiuchus 
star-forming region, the Orion star-forming region, and then two 
representative distances at which more massive star-forming regions 
start to be found.  (Alternatively, we can imagine that we are observing 
a star-forming region at a fixed distance of 160~pc and at a wavelength 
of 850~\um, but with telescopes of diameters 15~m, 5.4~m, 2.7~m, and 
1.3~m, respectively.)  To allow for careful scrutiny of the results, we 
always plot the same small (0.3$\times$0.3~pc) sub-section of the 
reference image.  The bottom row of figure~\ref{fig:resnoise} shows this 
sub-section as it would appear if observed from these four distances.

	 In nearby star-forming regions such as $\rho$~Oph and Orion, 
the typical size of presumed pre-protostellar cores is between 0.01~pc 
and 0.1~pc, with the average perhaps closer to 0.1~pc 
\citep{dj2000,dj2001,m01,jb2006,dj2006}.  Note that, although there is a 
lot of complex structure on the scale of pre-stellar cores in the 
simulated image at 0.16~kpc, all of this emission has been blended into 
a single object at 2~kpc.  Observing this region from 2~kpc at a single 
wavelength, it would be impossible to know whether this was a single 
massive clump 0.2~pc across or a collection of unresolved smaller 
clumps.

\begin{figure*}
\begin{center} 
\includegraphics[width=2\columnwidth,angle=270]{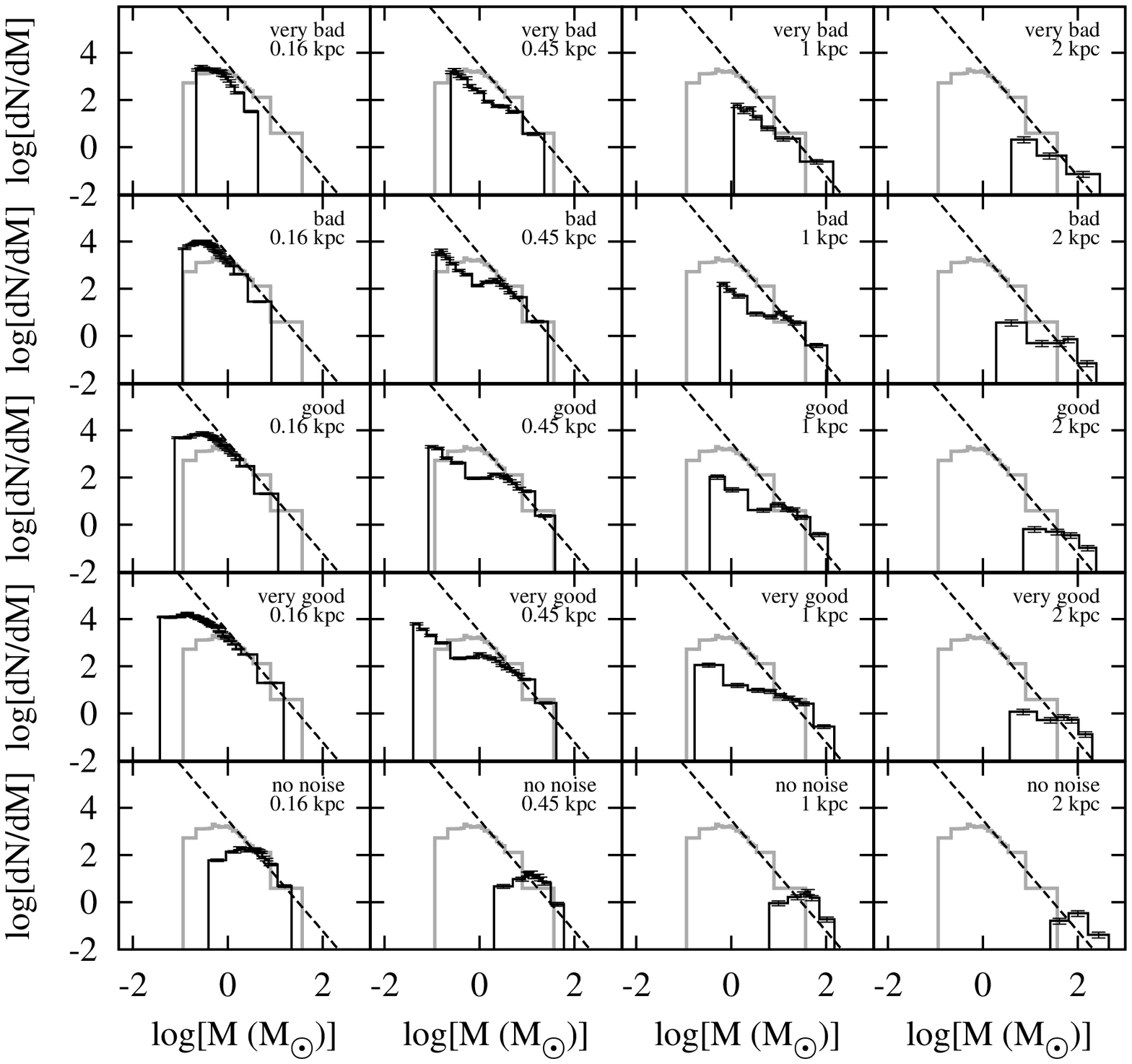} 
\caption{Differential clump mass functions (\emph{black lines}) 
extracted from versions of 
the reference image as convolved to the same resolutions and with the 
same levels of added noise as in Figure~\ref{fig:resnoise}.  Each DCMF 
is plotted over the DCMF of the clumps extracted from the reference 
image (\emph{grey line}) as shown in Figure~\ref{fig:refcmf}.  The 
dashed line is the Salpeter mass function, which is shown in the same 
position in all of the panels to guide the eye.
\label{fig:resnoisedmfs}} 
\end{center} 
\end{figure*}

\begin{figure*}
\begin{center} 
\includegraphics[width=2\columnwidth,angle=270]{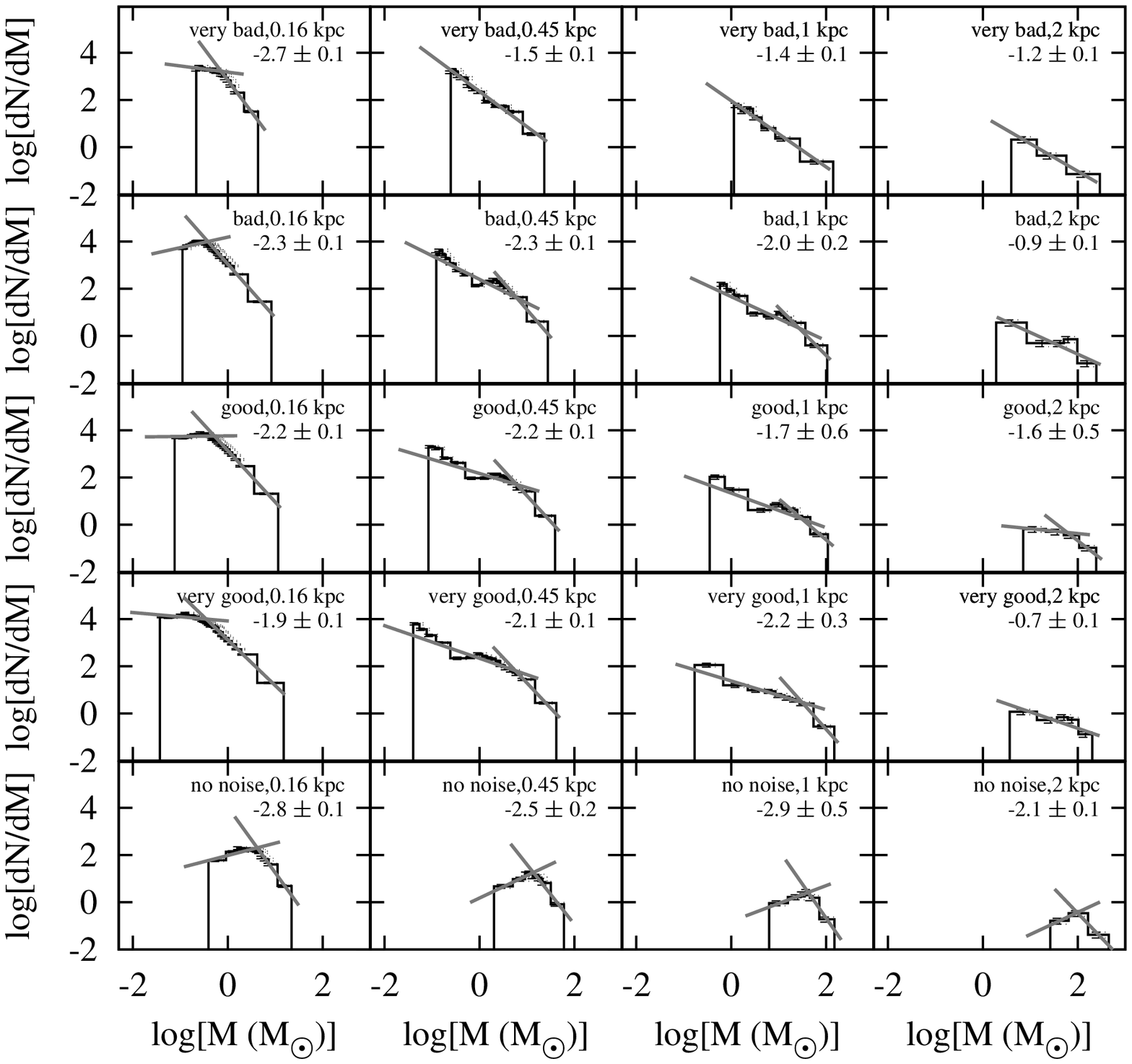} 
\caption{Differential clump mass functions (\emph{black lines}) as in 
Figure~\ref{fig:resnoisedmfs}, but with power-law fits (\emph{gray 
lines}).  Each mass function is fit with the best-fitting double power 
law or, where a double power law did not produce a good fit, a single 
power law.  The number indicated in the upper-right corner of each plot 
is the power-law index of either the single power law or the high-mass 
portion of the double power law.\label{fig:resnoisedmfsfits}}
\end{center} 
\end{figure*}

 	Now we turn to the mass functions.  The bottom row of 
Figure~\ref{fig:resnoisedmfs} shows the DCMFs for the clumps extracted 
from the convolved versions of the \emph{entire} reference image.  In 
each panel, the DCMF under study is compared to the reference DCMF from 
Figure~\ref{fig:refcmf} and the Salpeter power law.  Clearly, coarsening 
the angular resolution depletes low-mass clumps in this no-noise case.  
However, no progressive shallowing of the mass function is observed.  To 
make this point quantitatively, in Figure~\ref{fig:resnoisedmfsfits} we 
show the double power law which best fits each mass function.  As shown 
in the figure, the exponent of the high-mass end of the power-law does 
not decrease systematically with increasing distance.  At 0.16~kpc, it 
begins with a value of -2.8, somewhat lower than the nominal Salpeter 
value of -2.35, then drifts back and forth around a mean of -2.6 as the 
distance increases.

\begin{figure*}
\begin{center} 
\includegraphics[width=2\columnwidth,angle=270]{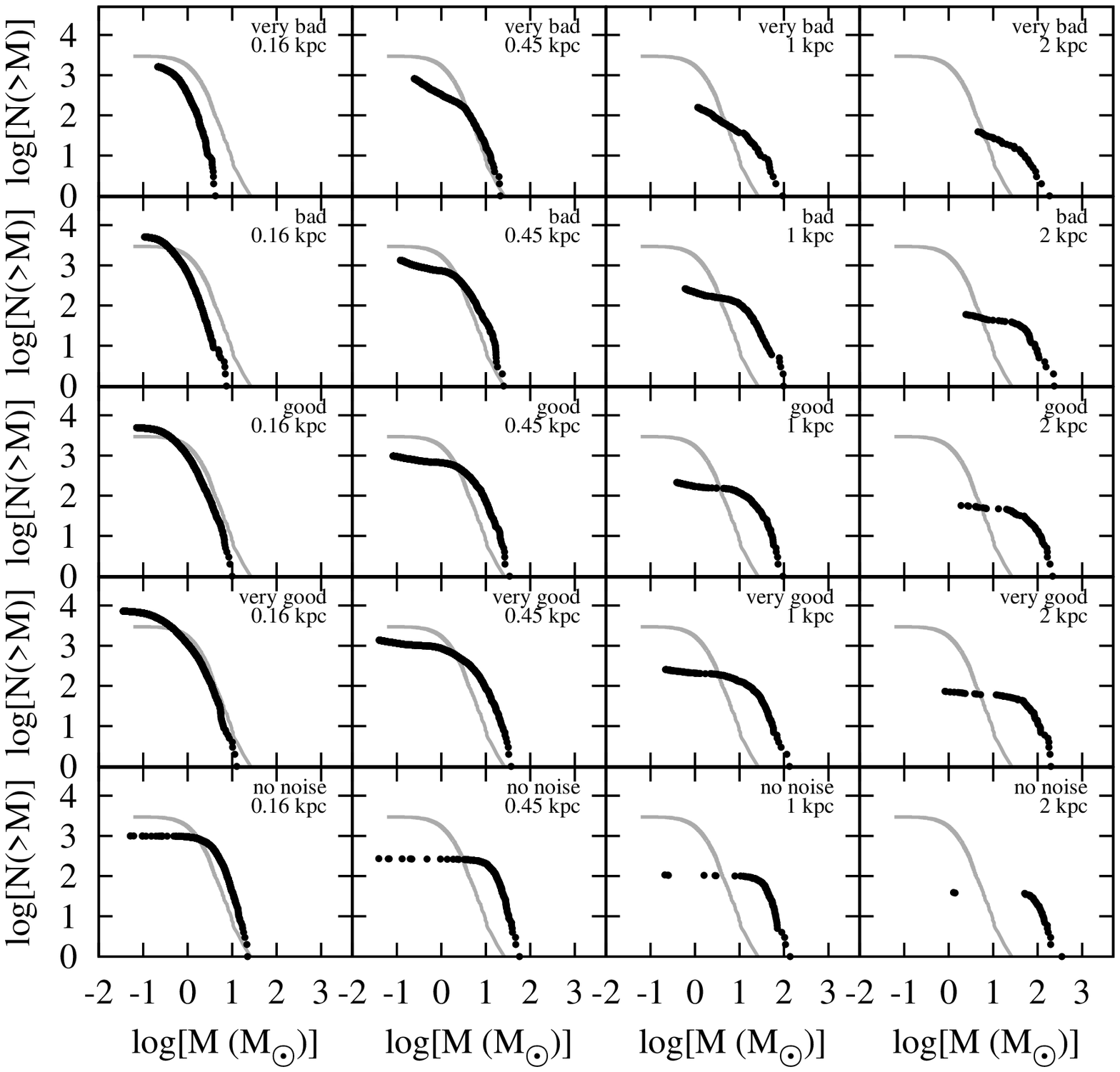} 
\caption{Cumulative clump mass functions (\emph{black dots}) 
extracted from versions of 
the reference image as convolved to the same resolutions and with the 
same levels of added noise as in Figure~\ref{fig:resnoise}.  Each CCMF 
is plotted over the CCMF of the clumps extracted from the reference 
image (\emph{grey line}) as shown in Figure~\ref{fig:refcmf}.  
\label{fig:resnoisecmfs}} 
\end{center} 
\end{figure*}

	Some of the changes in the mass function are obscured in the 
DCMF due to the decreasing total number of clumps.  For this reason, we 
also plot, in Figure~\ref{fig:resnoisecmfs} the CCMFs for each image.  
The CCMF does not suffer from ambiguities due to binning: it shows every 
single clump in the data set.  The Salpeter mass function is not drawn 
on these plots because, as described in \citet{rw06b}, the Salpeter mass 
function does not have a simple power-law form on the CCMF if it is 
assumed to have one on the DCMF.  Nevertheless, a similar result can be 
observed: in the no-noise case, there is no dramatic qualitative change 
in the shape of the CCMF as the resolution is coarsened.

	What should we make of the particular values of the power-law 
exponents obtained in the above fits?  In their meta-analysis of the 
stellar IMF, \citep{kroupa02} have shown that the slope of the high-mass 
end of the stellar IMF has a Salpeter-like mean of 2.36 with a standard 
deviation of 0.36.  Thus, we may say that clump mass functions with 
power-law slopes in the range 2.0 to 2.7 are consistent with the stellar 
IMF within the measurement uncertainties.  Within their uncertainties, 
all of the power-law fits to the no-noise DCMFs in 
Figure~\ref{fig:resnoisedmfsfits} would therefore seem to be consistent 
with the stellar IMF, although beyond the range usually interpreted to 
be so in studies of the clump mass function.

\subsection{Noise Effects}
\label{sec:noiseeff}

	In addition to finite angular resolution, real observations 
always have some level of noise, usually from both the sky and the 
instrument itself.  Adding noise suppresses the detection of fainter 
clumps and changes the fluxes of all clumps.  It must therefore affect 
the derived mass function.  The properties of the noise in an image may 
depend strongly on the technique used to produce the image.  For 
example, the scan-mapping technique common to most submillimeter 
cameras, including SCUBA \citep{scuba} and SCUBA2 on the JCMT and both 
SPIRE and PACS on Herschel, tends to produce images with fairly smooth 
noise across most of the image and a region of high noise around the 
edge of the image where coverage is incomplete.  Interferometer images, 
by contrast, typically have noise which is spatially very non-uniform.  
A single interferometric pointing results in noise which is low at the 
phase center but which climbs with distance from that center.  In an 
interferometric mosaic, the noise pattern can be much more complicated.

	For transparency and generality, we have adopted a simple 
prescription for adding noise to images.  This technique most closely 
approximates the type of noise found in scan maps, which we argue are 
the type most commonly used in measuring the clump mass function and 
best suited for that purpose. We add Gaussian random noise on a 
per-pixel basis to the reference image and then convolve it with the 
beam, scaling the amplitude of the added noise appropriately so that it 
has the desired amplitude after convolution with the selected beam.  
This ensures that the noise in the image is correlated on angular 
scales 
matching the beam, as it is in real observations.  To mimic a wide range 
of integration times and weather grades, we chose four representative 
noise levels which we label very good, good, bad, and very bad.  Our 
``good'' noise value, $\sigma_{\rm good}$, approximates the actual noise 
obtained in a typical 10\arcmin$\times$10\arcmin~SCUBA 850~\um~scan map 
over 10 hours of integration in grade 2 weather (roughly 0.03 Jy 
beam$^{-1}$).  The noise levels of the four grades differ by factors of 
2, so that $\sigma_{\rm very\ good} = \sigma_{\rm good}/2 = \sigma_{\rm 
bad}/4 = \sigma_{\rm very\ bad}/8$.  Modern instruments comparable to 
SCUBA, such as SCUBA2 and SPIRE, will typically produce images with less 
noise than our ``good'' level, but many archival observations will be 
closer to the ``bad'' noise level.

	The noise-added versions of the reference image subsections 
discussed earlier are also included in Figure~\ref{fig:resnoise}.  Each 
row of the figure shows a different noise value.  Scanning up a column 
shows the same subsection of the simulated region at a constant angular 
resolution but a progressively higher level of noise.  Similarly, 
Figures~\ref{fig:resnoisedmfs}, \ref{fig:resnoisedmfsfits}, and 
\ref{fig:resnoisecmfs} show the DCMFs 
and CCMFs, respectively, of the reference image at each noise level and 
angular resolution.

	Scanning up any given column in Figure~\ref{fig:resnoisedmfs} 
reveals that the effects of increased noise are most pronounced at low 
clump masses and near distances.  Note that the addition of even a small 
amount of noise, as in the 0.16~kpc, very good noise image, increases 
the number of low mass clumps in the mass function.  This occurs in part 
because the addition of noise raises some previously undetected peaks 
above the detection 
threshold, but also because noise can break apart more massive clumps, 
tricking \clfind\ into thinking that they are multiple clumps of lower 
mass.  This effect could be minimized in future by using 
multi-wavelength data sets in which each clump is observed independently 
more than once.  Herschel will be very helpful in this regard.

	However, note also that, at least in nearby regions, the 
addition of even substantial amounts of noise does not make the measured 
mass function conclusively non-Salpeter.  This is evident from the fits 
in Figure~\ref{fig:resnoisedmfsfits} in which we see that most of the 
0.16~kpc and 0.45~kpc mass functions are well fit by power laws whose 
exponents fall within the IMF-like range described in 
\S\ref{sec:reseffects}.  Even at the higher distances of 1~kpc and 
2~kpc, some of the mass functions retain this IMF-like shape.  Referring 
back to Figure~\ref{fig:resnoise}, we should be surprised by this 
result: the bad-noise images definitely do not all show the same 
population of objects, yet all but the 2~kpc versions are well fit by 
Salpeter-like power laws.  The potentially worrisome conclusion is that 
\emph{even populations of clumps which are certainly not representative 
of pre-stellar objects may present Salpeter-like mass functions}.  If 
the clumps in the 0.16~kpc bad noise simulation are representative of 
pre-stellar clumps, surely the ones in the 1~kpc bad noise simulation 
are not, yet they are both well fit by mass functions which are 
compatible with the stellar IMF.

\subsection{Chopping, Interferometry, and `Spatial Filtering'}
\label{sec:spatfilt}

	Most measurements of the clump mass function made to date have 
been made from spatially filtered images.  By ``spatially filtered 
images'', we mean those which do not reproduce the emission from the 
source faithfully on all spatial scales.  Instead, the source brightness 
is usually sampled on some finite set of spatial scales and then its 
image reconstructed using a Fourier-type method.  Most often, this 
spatial filtering takes the form of chopping 
\citep{man98,dj2000,dj2001,m01,rw05,rw06a} but it can also take the form 
of interferometry (e.g. \citealt{ts98})

	In chopped images, some of the flux from the source is lost.  A 
full discussion of chopping and the common techniques for reconstructing 
images from chopped data can be found in \citet{e95,jlh98}, and 
\citet{scuba}.  Using a single chop throw is equivalent to multiplying 
the telescope's spatial frequency response by a sine function, meaning 
that emission at all but a few spatial scales is attenuated to some 
degree \citep{e95}.  A better strategy is to combine data sets made 
using several chop throws which have a small greatest common factor.  In 
this way, the first null of the attenuating sine function can be made to 
correspond to angular scales much larger than any of interest in the 
image.  Thus, the actual emission from the source can be reconstructed 
with high fidelity (or perfect fidelity, in the limit of no noise).  
Clump mass functions produced from images acquired in this way should be 
relatively unaffected by the spatial filtering.

	Interferometric observations are typically more strongly 
filtered than chopped images, so the effect on the clump mass function 
of this filtering may be more pronounced and harder to predict.  Like 
chopped images, interferometric images only include emission from the 
source on certain ranges of spatial scales.  Optimizing for high angular 
resolution typically sacrifices emission on large spatial scales.
Modern interferometers with many elements attempt to maximize 
both angular resolution and sensitivity to a broad range of spatial 
scales.  For example, ALMA will incorporate the smaller Atacama Compact 
Array (ACA), which will improve its sensitivity to emission on large 
spatial scales. 
	
\begin{figure*} 
\begin{center} 
\includegraphics[width=1.25\columnwidth,angle=270]{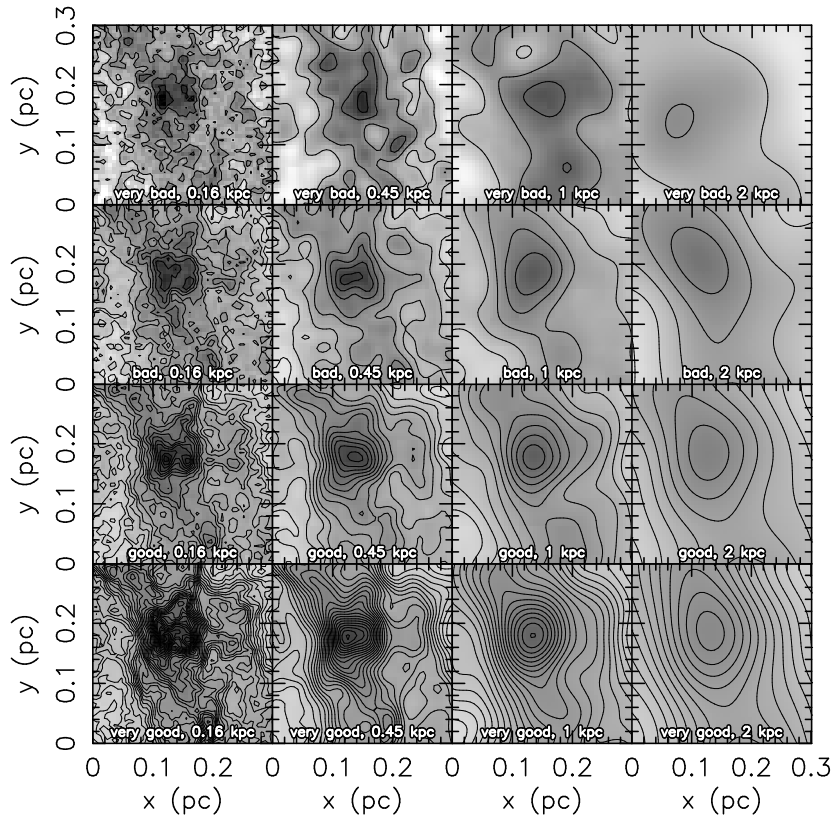} 
\caption{Representations of the chopped versions of the reference image 
as seen at four different angular resolutions (\emph{x axis}) and four 
different noise levels (\emph{y axis}).  The areas covered and the 
contour levels used are identical to those in Figure~\ref{fig:resnoise}.  
Note that there is no noise-free version of each image because, in the 
limit of no noise, a chopped image is identical to the original.
\label{fig:chopimgs}}
\end{center} 
\end{figure*}

\begin{figure*}
\begin{center} 
\includegraphics[width=2\columnwidth,angle=270]{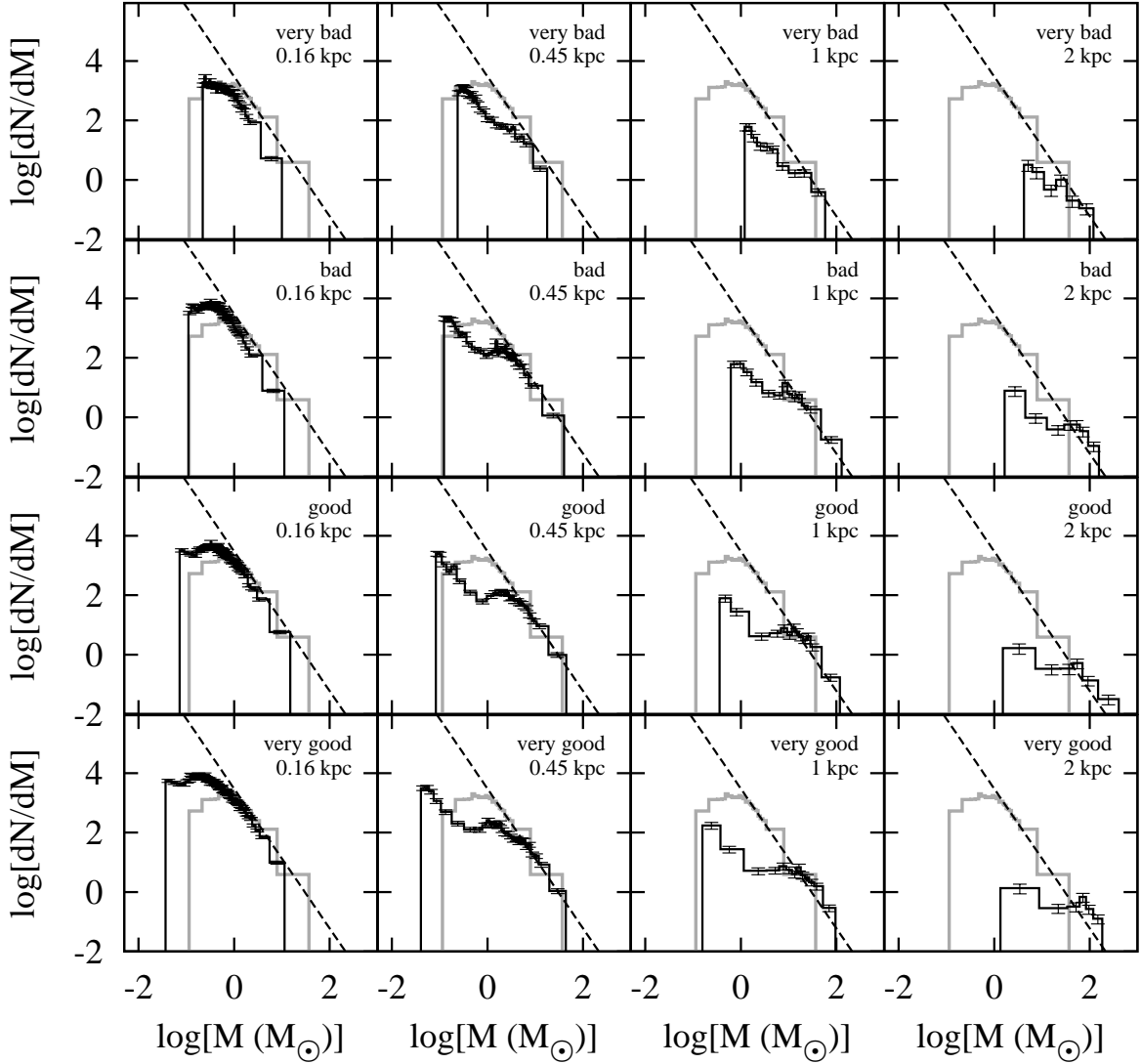} 
\caption{Differential clump mass functions (\emph{black lines}) 
extracted from the chopped versions of the reference image.  The 
angular resolutions and noise levels are as described in 
Figure~\ref{fig:resnoisedmfs}, but for the lack of the no-noise 
versions.   Each DCMF is plotted over the DCMF of the 
clumps extracted from the reference image (\emph{grey line}) as shown in 
Figure~\ref{fig:refcmf}.  The dashed line is the Salpeter mass function, 
which is shown in the same position in all of the panels to guide the 
eye. 
\label{fig:chopmfs}}
\end{center} 
\end{figure*}

\begin{figure*}
\begin{center} 
\includegraphics[width=2\columnwidth,angle=270]{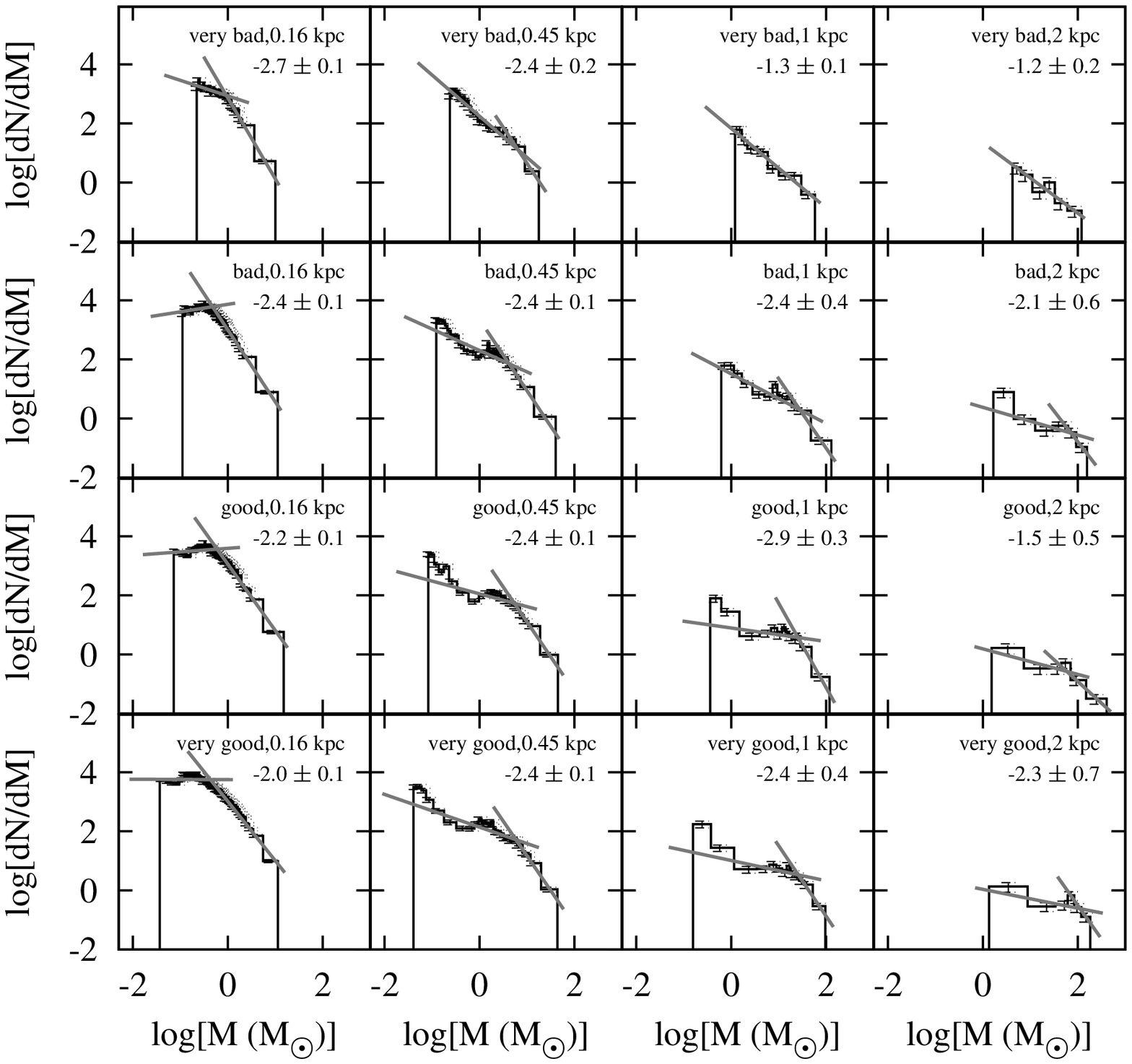} 
\caption{Differential clump mass functions (\emph{black lines}) as in 
Figure~\ref{fig:chopmfs}, but with power-law fits (\emph{gray 
lines}).  Each mass function is fit with the best-fitting double power 
law or, where a double power law did not produce a good fit, a single 
power law.  The number indicated in the upper-right corner of each plot 
is the power-law index of either the single power law or the high-mass 
portion of the double power law.\label{fig:chopmfsfits}}
\end{center} 
\end{figure*}

	Rather than attempting to simulate many specific cases of 
spatial filtering, we have chosen two generic representations: a common 
type of chopping and a moderate form of interferometric filtering. In 
the first, we simulate a chopped image taken with three chop throws.  We 
chose the recommended minimum of three chop throws with no large common 
factor , in 
this case 30\arcsec, 44\arcsec, and 68\arcsec.  Figure~\ref{fig:chopimgs}
 shows the same subsections of the reference image as 
Figure~\ref{fig:resnoise}, but with chopping applied.  The corresponding 
clump mass functions are compared with the reference mass function in 
Figure~\ref{fig:chopmfs} and shown with best-fitting power laws in 
Figure~\ref{fig:chopmfsfits}.  Note that there is no noise-free case 
shown in these figures as, in the absence of the noise, a chopped image 
is identical to the original so that the no-noise results in this case 
would be the same as in the bottom row of Figure~\ref{fig:resnoisedmfs}.

	Comparison of Figures~\ref{fig:chopmfs} and 
\ref{fig:resnoisedmfs} shows that, as expected, the clump mass functions 
do not change much due to chopping, as long as a sufficient number of 
chop throws with a small greatest common factor are used (three 
suffices).

	The same cannot be said of mass functions measured from more 
heavily filtered images, such as those produced through interferometry.  
We simulate the effects of interferometric spatial filtering in a 
general way not tied to any particular instrument.  We choose to 
suppress emission on scales larger than about 2\arcmin.  To do so, we 
multiply the Fourier transform of the image by a circularly symmetric 
Gaussian whose full-width at half maximum corresponds to angular scales 
of 2\arcmin in the image plane.  Our choice of a Gaussian taper with 
this width is somewhat arbitrary: different tapering functions with 
different widths would correspond to different interferometer 
configurations.  We performed many such manipulations of the image but, 
as the essential results do not change, we present only this set for 
brevity.

\begin{figure*} 
\begin{center} 
\includegraphics[width=1.5\columnwidth,angle=270]{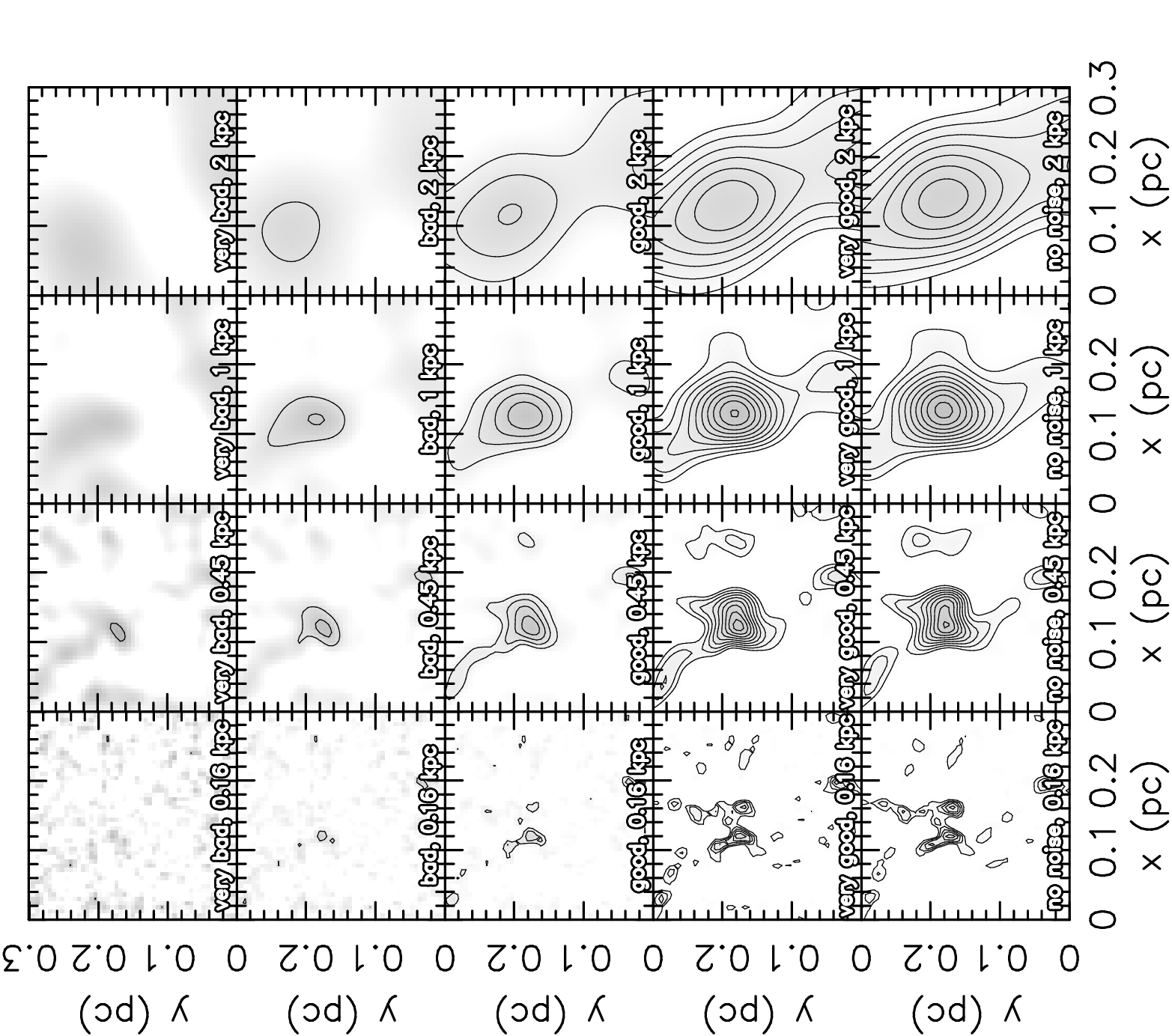} 
\caption{Representations of the versions of the reference image 
as seen at four different angular resolutions (\emph{x axis}) and four 
different noise levels (\emph{y axis}), but with emission on scales 
larger than 2\arcmin~strongly suppressed to mimic interferometric 
observations.  The areas 
covered and the 
contour levels used are identical to those in Figure~\ref{fig:resnoise}.  
\label{fig:gausimgs}}
\end{center} 
\end{figure*}

\begin{figure*}
\begin{center} 
\includegraphics[width=2\columnwidth,angle=270]{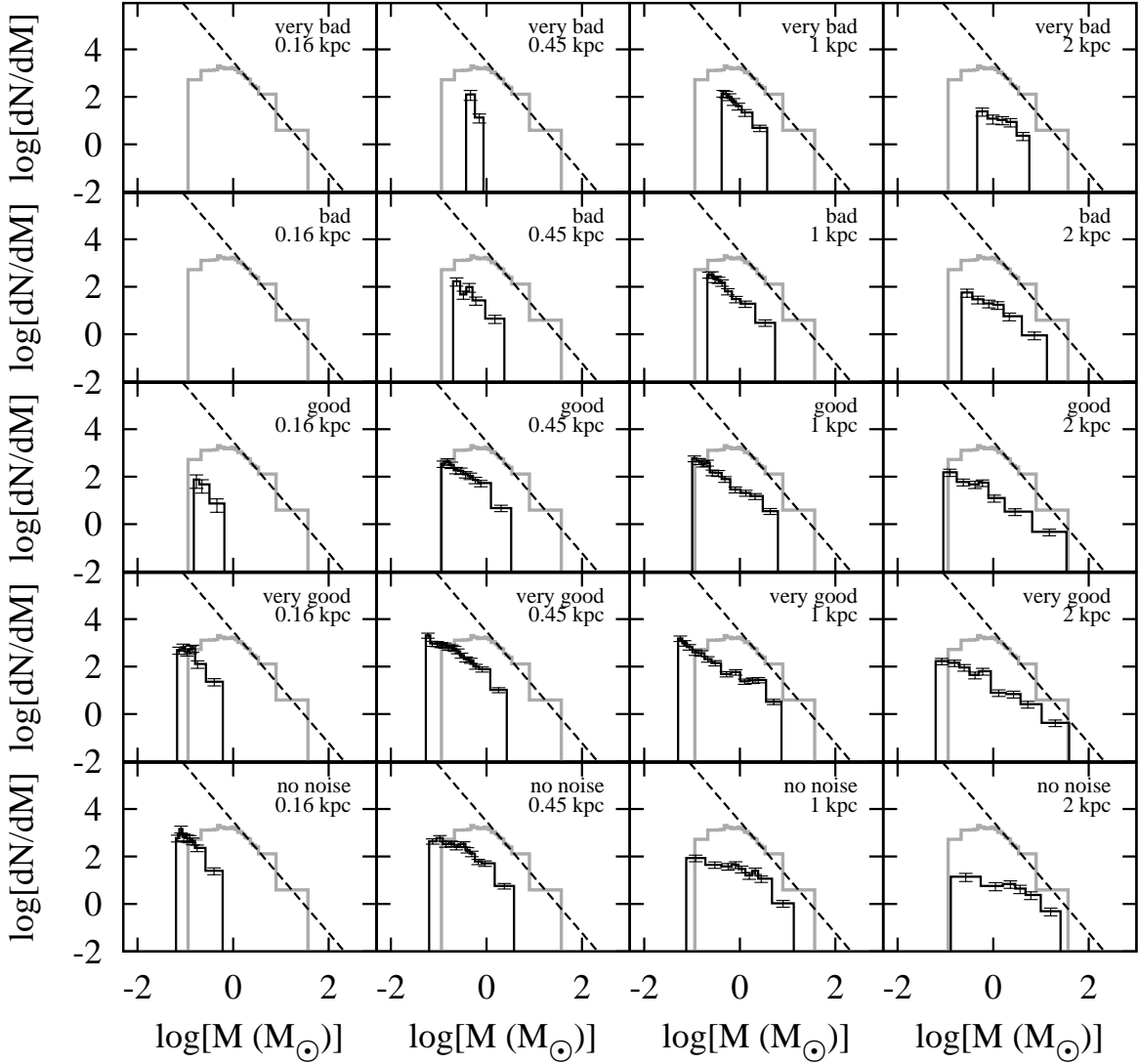} 
\caption{Differential clump mass functions (\emph{black lines}) 
extracted from the versions of the reference image in which emission on 
spatial scales larger than 2\arcmin~has been strongly suppressed.  The 
angular resolutions and noise levels are as described in 
Figure~\ref{fig:resnoisedmfs}.  Each DCMF is plotted over the DCMF of 
the 
clumps extracted from the reference image (\emph{grey line}) as shown in 
Figure~\ref{fig:refcmf}.  The dashed line is the Salpeter mass function, 
which is shown in the same position in all of the panels to guide the 
eye. 
\label{fig:gausmfs}}
\end{center} 
\end{figure*}

\begin{figure*}
\begin{center} 
\includegraphics[width=2\columnwidth,angle=270]{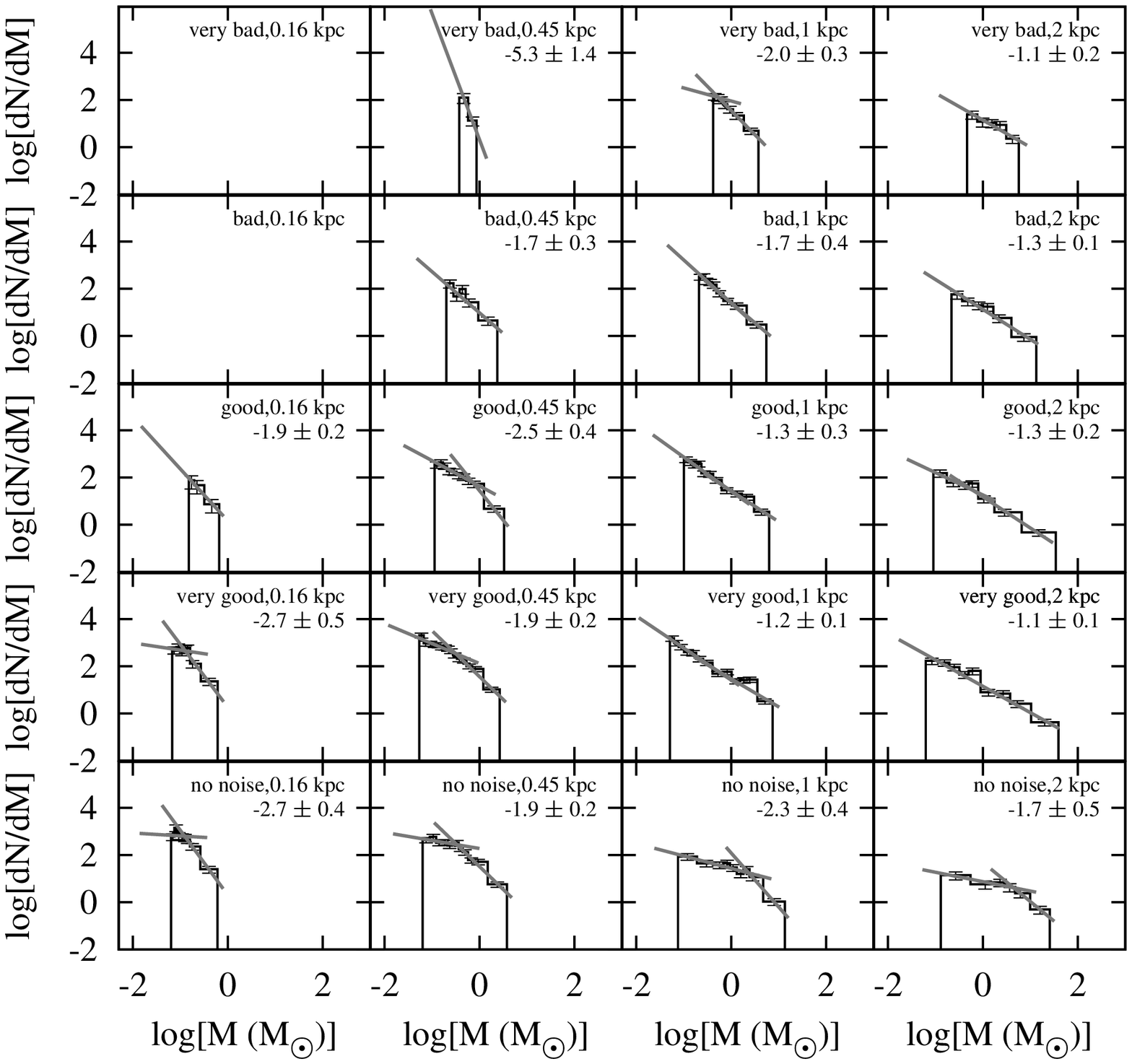} 
\caption{Differential clump mass functions (\emph{black lines}) as in 
Figure~\ref{fig:gausmfs}, but with power-law fits (\emph{gray 
lines}).  Each mass function is fit with the best-fitting double power 
law or, where a double power law did not produce a good fit, a single 
power law.  The number indicated in the upper-right corner of each plot 
is the power-law index of either the single power law or the high-mass 
portion of the double power law.\label{fig:gausmfsfits}}
\end{center} 
\end{figure*}

	As before, Figure~\ref{fig:gausimgs} shows the standard 
subsection of the reference image, now filtered.  
Figure~\ref{fig:gausmfs} shows the corresponding DCMFs compared to the 
reference mass DCMF and Figure~\ref{fig:gausmfsfits} shows the DCMFs 
fitted with double power laws.  

	These heavily spatially filtered images produce a marked change 
in the clump mass function.  They demonstrate how difficult it can be to 
predict the effect on the clump mass function of different types of 
image acquisition techniques.  In the 0.16~kpc maps with bad and very 
bad noise, all of the clumps have fallen below the detection threshold.  
This again reminds us that the threshold for detection of clumps is a 
surface brightness, not an absolute flux.  Because these two simulated 
regions are nearby, the emission from their clumps occurs on relatively 
large angular scales, so they are filtered out by the simulated 
interferometry.  Thus, as Figure~\ref{fig:gausmfs} shows, the DCMFs of 
the nearer regions are actually \emph{less} similar to that of the 
reference DCMF than are those of the more distant regions.  Measurements 
of the clump mass function with interferometers such as ALMA must take 
this effect into account.

	The shape of the mass function is strongly affected by 
interferometry.  In several cases, double power law fits now simply 
converge to single power laws; in those cases, we show the single power 
law and its exponent in Figure~\ref{fig:gausmfsfits}.  Interferometry 
does us the favor of removing distracting emission from the diffuse 
background which is probably not directly involved in star formation.  
Consulting Figure~\ref{fig:gausimgs}, we can see that the clumps are 
much more visible than in Figure~\ref{fig:resnoise}.  Perhaps as a 
result, the mass functions in Figure~\ref{fig:gausimgs} are now 
consistently shallower than the Salpeter IMF.  Yet still there are cases 
where, although the images themselves show that the population of 
clumps has changed substantially due to noise, resolution, and 
filtering effects, the mass function is well-fit by a Salpeter-like 
power law.  Again, it appears that even when the objects from which the 
clump mass function is made are not themselves the precursors of 
individual stars, the mass function may appear Salpeter-like.

\subsection{Implications for Measurements of the Clump Mass Function}
\label{sec:imps}

	What are the consequences of these results for measurements of 
the clump mass function?  In attempts thus far to measure the clump 
mass function and compare it to the stellar IMF, the comparison has 
often been framed in this form: ``Does this mass function look like a 
Salpeter 
power-law?''  However, we have shown that there are many cases in which 
a population of objects--poorly resolved, noisy blobs derived from 
observations of star-forming regions--may yield Salpeter-like mass 
functions.  Hence, we suggest that a better question would 
be ``Do I have reason to believe that my observations would \emph{not} 
yield a Salpeter-like mass function?''  If the observations cannot be 
expected \emph{a priori} to definitively discriminate between 
Salpeter and non-Salpeter forms, then any resulting appearance of a 
Salpeter-like clump mass function should not be over-interpreted.  

	Our best prospects for measuring the clump mass function 
accurately rest with single-dish telescopes, where spatial filtering can 
be minimized and sensitivity maximized over a large field of view.  
Herschel and SCUBA2 promise to be a powerful combination 
in this regard because both have achieved unprecedented sensitivity in 
their respective wavebands and they have complementary resolving power: 
where Herschel is unable to resolve individual pre-stellar cores at long 
wavelengths, SCUBA2 will provide the required resolving power.  

	In interpreting clump mass functions produced from single-dish 
observations, readers should remember that having a Salpeter-like mass 
function is not conclusive evidence that a population of clumps 
represent 
individual pre-stellar cores.  Rather, the argument should be approached 
from the opposite direction: do these objects that I know are 
pre-stellar cores (by other lines of evidence), have a Salpeter-like 
mass function?  

\subsection{Characteristic Scales of the Mass Function and the Star 
Formation Efficiency}

	Some authors have noted that the characteristic mass scales of 
the stellar IMF and clump mass functions seem to match, modulo some star 
formation efficiency (e.g. \citealt{dj2000,dj2001,m01,m07}).  The idea 
is that one can obtain the stellar IMF from the clump mass function by 
multiplying the masses of all of the clumps by some star formation 
efficiency less than unity.  To derive this efficiency, one can compute 
the ratio of certain characteristic masses measured from both mass 
functions.  The clump mass function has two basic characteristic masses: 
the ``peak'' or ``turnover'' mass, which is simply the mass at which the 
mass function peaks, and the ``break mass'', which is the break point 
between the two power laws when the mass function is fit with a double 
power law.  The break mass is equal to or greater than the peak mass.  
One can compute a star formation efficiency from either value.  For 
example, if the break mass in the stellar IMF were 0.3~\msun\ and that 
in the mass function of pre-stellar cores were 3~\msun\, one might 
conclude that the star formation efficiency was about 10\%.

\begin{figure} 
\begin{center} 
\includegraphics[width=\columnwidth,angle=270]{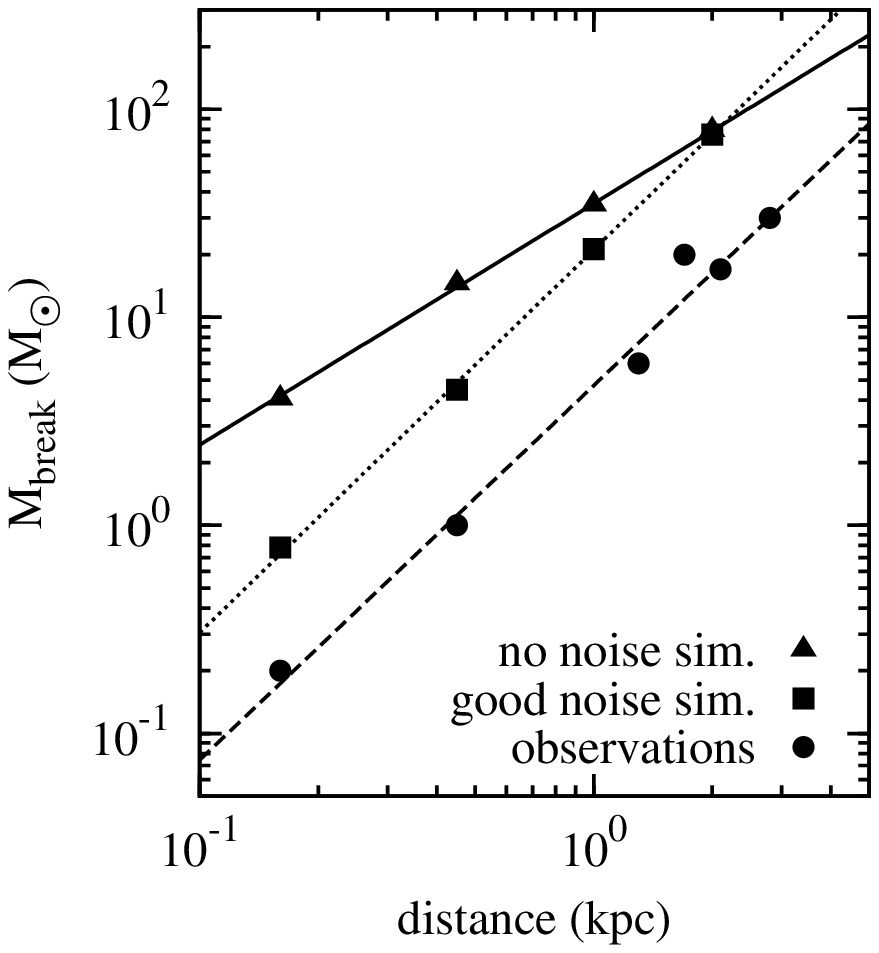} 
\caption{Break mass versus distance to various observed and simulated 
star-forming regions.  The break mass is a parameter of a double power 
law fit to a clump mass function and represents the mass at which the 
two power laws intersect.  Break masses from the observational data are 
shown with filled circles, while those from the simulations with no 
noise and good noise are shown with filled triangles and squares, 
respectively.  The lines of best fit shown are power laws in distance 
with exponents of 1.2 for the noise-free simulations (\emph{solid line}) 
and 1.8 for both the simulations with good noise (\emph{dotted line}) 
and the 
observations (\emph{dashed line}). \label{fig:breakmass}} 
\end{center} 
\end{figure}

	However, this interpretation of these results may also be too 
simplistic.  Scanning across any row in 
figure~\ref{fig:resnoisedmfsfits} demonstrates that, as distance 
increases, both the peak and break masses also increase.  In effect, the 
whole mass function shifts to higher masses.  To make this point 
explicit, we plot in Figure~\ref{fig:breakmass} the fitted DCMF break 
mass versus the distance to several star-forming regions, both simulated 
and real.  The real data come from fits to the mass functions of the 
following star-forming regions, in order of increasing distance: 
$\rho$~Oph \citep{dj2000,man98}, Orion~B \citep{dj2001,m01}, M8 
\citep{tot02}, Cygnus~X \citep{m07} (using data read from their 
published graphs), M17 \citep{rw06a}, and NGC~7538 \citep{rw05}.  For M8 
and M17, we have used the revised distance estimates of 1.3~kpc 
\citep{tot08} and 2.1~kpc \citep{cm08}.  

	Fitting the simulated data with no noise and good noise, we find 
that the break mass scales with distance as $d^{1.2}$ and $d^{1.8}$, 
respectively.  Fitting the data derived from actual observations, we 
find that the break mass scales with distance again as $d^{1.8}$.  The 
simulations produce consistently higher break masses than do the 
observations.  We cannot account quantitatively for this result yet, but 
we suspect it results from the much larger areal coverage (and hence 
larger number of clumps) in the simulations than in the observations.  
Having more clumps in the sample may increase the statistical weight of 
the high-mass end of the mass function, where the statistics within the 
observations are usually poor (because higher mass clumps are rarer).

	Unlike the simulated images, which differ only in the distance 
to the simulated region, the observational data sets differ widely in 
the telescope used to acquire the data, the wavelength at which the data 
were acquired, and the clump-finding algorithm used to extract the 
clumps.  As we mentioned in \S\ref{sec:define}, this supports the 
argument that the properties of derived clump mass functions do not 
depend strongly on the choice of clump-finding algorithm.

	That the break mass should scale with the distance to the region 
being observed is exactly what one would expect if the break mass were 
not a property of the ensemble of clumps themselves, but a function of 
the angular resolution at which they are observed.  If the break mass 
reflected, say, the local Jeans mass or some property of turbulence, it 
ought not to scale with the distance to the observed region if 
the all regions were observed with sufficiently high angular 
resolution.  

	This scaling of the break mass with distance is already implicit 
in other results in the literature.  For example, \citet{m07} showed 
that, in the massive star-forming complex Cygnus~X, which lies at a 
distance of 1.7~kpc, the typical volume-averaged density of a clump is 
about an order of magnitude lower than that of the clumps in $\rho$~Oph, 
which lies at only 0.16~kpc.  This could as easily be a resolution 
effect as a physical one.

	If the characteristic masses of the clump mass functions 
observed to date really are set more by the distances to the regions 
observed than by the physics of the material within those regions, then 
their use in deriving star formation efficiencies must be questioned.  
Herschel and SCUBA2 will both offer observations with 
improved angular resolution and sensitivity, allowing us to test the 
robustness of this distance scaling.  

\section{The Lognormal Mass Function}

	We showed in \S\ref{sec:noiseeff} that, even when a population 
of clumps has been distorted beyond recognition by noise and coarse 
angular resolution, its mass function may still be consistent with the 
stellar IMF within the uncertainties.  We believe that this behavior 
of the clump mass function reflects the deeper origins of the shape of 
the clump and stellar mass functions.

	The stellar IMF is frequently described as a Salpeter-like 
power-law.  This interpretation is somewhat outdated; it reflects the 
limited range of stellar masses included in Salpeter's original stellar 
IMF.  More recent measurements of the stellar IMF show that, when it is 
extended to low stellar masses, well below the turnover mass, it adopts 
a lognormal form \citep{chabrier03}, sometimes approximated by four or 
five power-law segments of which the Salpeter power-law is but one.  The 
Salpeter power law appears merely to be a good approximation to the 
lognormal IMF over a restricted range of stellar masses, perhaps 
1--10~\msun.

	Several authors have shown that a lognormal stellar IMF arises 
naturally as a consequence of the Central Limit Theorem of calculus 
\citep{larson73,z84,af96}.  When a sufficiently large number of 
independent physical processes or variables act together to produce the 
stellar IMF, it naturally tends towards a lognormal form.  The larger 
the number of independent variables used in models of the origin of the 
IMF, the more closely the IMF approaches the lognormal form 
\citep{af96}.  

\begin{figure*}
\begin{center} 
\includegraphics[width=2\columnwidth,angle=270]{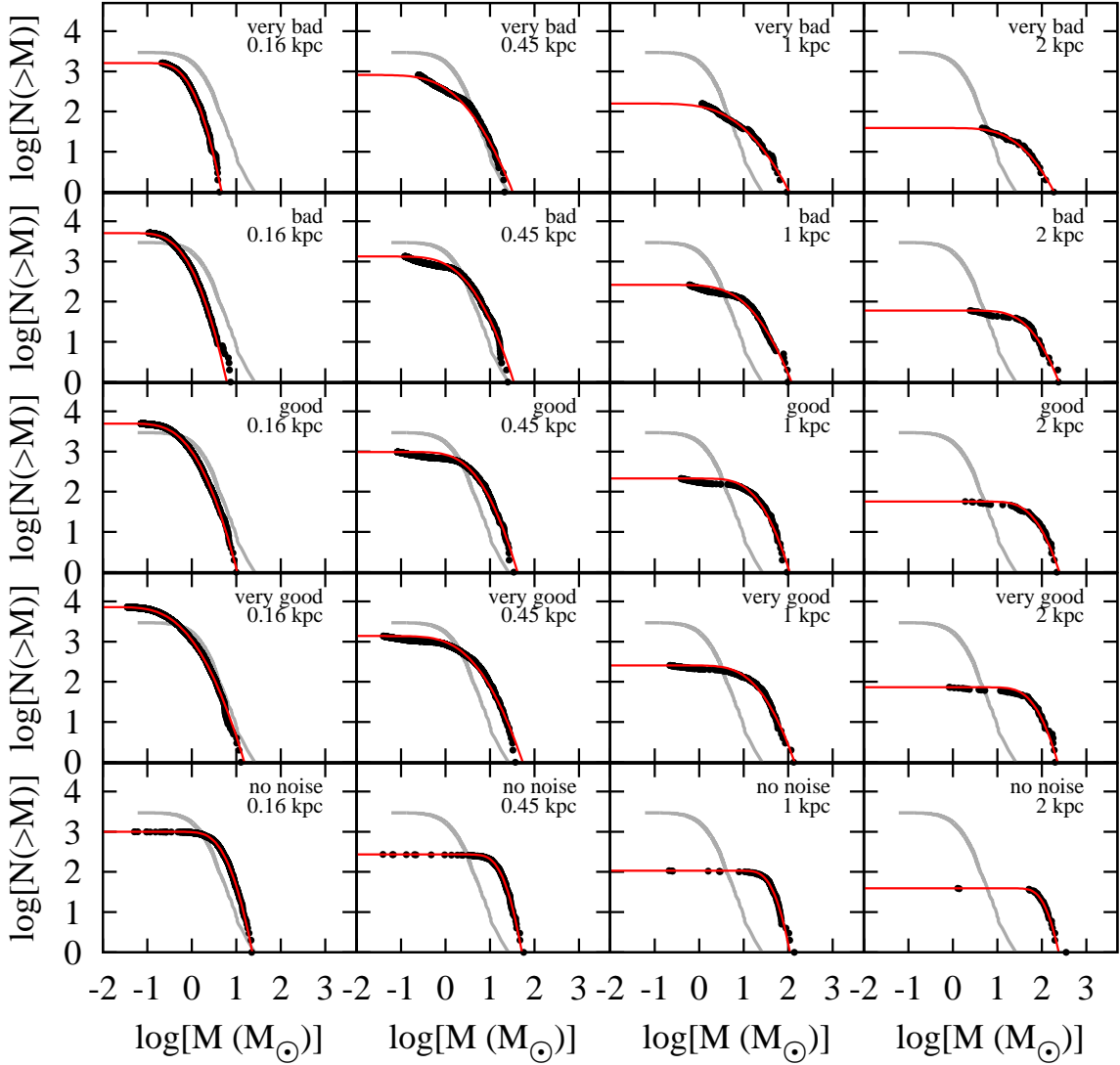} 
\caption{Cumulative clump mass functions (\emph{black dots}) identical 
to 
those in 
Figure~\ref{fig:resnoisecmfs}, including the mass function of the 
reference image (\emph{grey line}), but adding a fitted error function 
(\emph{red line}).  The 
error function is the CCMF equivalent of a lognormal DCMF.
\label{fig:logncmfs}}
\end{center} 
\end{figure*}

	The same reasoning applies to the clump mass function for two 
reasons: first, a large number of independent factors must act to set 
the distribution of clump masses and, second, it is that distribution of 
clump masses which must ultimately give rise to the IMF.  \citet{rw06b} 
showed that clump mass functions measured from dust continuum maps are 
typically well fit by lognormal functions.  This result holds true 
despite the large variety of different methods of acquiring the data and 
extracting the clumps used in producing the various mass functions.  In 
Figure~\ref{fig:logncmfs}, we show that the mass functions first shown 
in Figure~\ref{fig:resnoisedmfs} are all well fit by lognormal 
distributions.  This is easier to see when plotting CCMFs, which show 
every clump in the sample.  A lognormal DCMF corresponds to an error 
function.  We believe that the high quality of the lognormal fits to the 
simulated clump mass functions suggest that they, too, are being biased 
toward this form by the cumulative action of a large number of 
independent processes.  These factors need not be physical processes, as 
assumed in models of the origin of the stellar IMF.  The independent 
processes might equally well include those discussed in 
\S\ref{sec:define}, namely things like the addition of random amounts of 
noise to each clump, random errors in the assignment of clump boundaries 
by clump-finding algorithms, and the perturbations to each clump's mass 
caused by convolution with other clumps and superposition along the line 
of sight.  Indeed, as the plots in Figure~\ref{fig:logncmfs} show, the 
combined effects of large amounts of noise and very coarse angular 
resolution do not ruin the lognormal shape of the mass function.

	Our reference mass function already has a lognormal form.  The 
only post-processing required to construct that mass function was the 
use of a clump-finding algorithm; excessive noise and coarse angular 
resolution are not factors.  Adding noise and coarsening the resolution 
of the simulation appears to change the width and normalization of the 
lognormal clump mass function, but not to make it any less lognormal.  
This is the behavior one expects if the Central Limit Theorem is 
setting the clump mass function.  However, it raises the prospect that 
it may be very difficult indeed to measure a clump mass function which 
is \emph{not} lognormal.  If all clump mass functions appear lognormal, 
it may be difficult to distinguish those whose width and normalization 
were set by the physics of star formation and those whose 
characteristics were set by our data acquisition and analysis 
techniques.

\section{Conclusions}

	We have investigated the effects on the derived clump mass 
function of image noise, image angular resolution, and two kinds of 
spatial filtering.  We have found that adding noise to an image and 
coarsening its resolution to the point where the objects in the image 
are clearly no longer the precursors of individual stars frequently does 
not cause its mass function to become incompatible with the Salpeter 
stellar IMF.  When the simulated mass functions are fit with power laws, 
the distribution of the power law exponents caused by noise and degraded 
resolution mirrors the distribution of measured exponents in the stellar 
IMF.  The clump mass function only deviates conclusively from the 
Salpeter form when it is derived from heavily spatially filtered 
observations.

	Following other authors \citep{larson73,z84,af96}, we have 
suggested that the clump mass function has a lognormal form due to the 
cumulative action of many independent processes in determining the mass 
of any given clump.  We have extended the set of such possible processes 
to encompass not only physical processes occurring in star-forming 
regions, but the processes of data acquisition and analysis.  The 
cumulative action of factors such as turbulence, temperature variations, 
radiative effects, numerous uncertainties in our conversion of flux to 
mass, our clump-finding algorithms, image noise, source blending, and 
spatial filtering may \emph{ensure} that clump mass functions always 
appear lognormal.  The Salpeter-like appearance of their high-mass ends 
may simply reflect this lognormal form.

	Collectively, these results suggest that we ought to adopt a 
skeptical approach when interpreting the clump mass function.  We cannot 
conclude that, because the mass function of a set of clumps has a 
Salpeter-like form, those clumps represent the precursors of individual 
stars.  We \emph{may} be able to draw this conclusion in the very 
limited set of cases in which our observations have very little noise 
and sufficient resolution to distinguish individual pre-stellar cores.  
Dust continuum maps made with Herschel's PACS instrument have a lot of 
promise in this regard because they have unparalleled sensitivity and 
resolution.  However, we will demonstrate in a forthcoming paper that 
even clump mass functions derived from PACS maps can show a convincing 
Salpeter-like form in cases where we do not believe the constituent 
clumps to be pre-stellar cores.

	The study of the origin of the stellar IMF is important to many 
areas of astronomy.  It is worth very careful scrutiny.  We suggest that 
the best measurements of the clump mass function with the current 
generation of instruments will come from a combination of PACS and 
SCUBA2 data.  High-sensitivity, multi-wavelength observations at high 
spatial resolution will allow us to reduce the effects of many of the 
uncertainties discussed in this paper.  Using the multi-wavelength data, 
we will be able to better constrain the temperatures and dust opacities 
of the clumps, improving our estimates of their masses.  Using Spitzer 
and PACS data, we will be able to do a better job of filtering out cores 
which are already forming stars.  Follow-up observations to obtain 
high-resolution molecular line data will further allow for the exclusion 
of clumps which are not gravitationally bound, as well as limited 
deconvolution of the emission along the line of sight.  

	The comparison of Herschel and SCUBA2 clump catalogs for 
matching regions will be highly instructive.  Such a comparison would 
allow us to assess, in a quantitative and statistically significant way, 
whether our measurements of the masses of individual clumps are robust.  
If, for example, both SCUBA2 and Herschel generate Salpeter-like mass 
functions, but with very different masses for each individual clump, 
this will be good evidence for our argument that the shape of the 
clump mass function can be set by non-physical means.  We will be very 
reassured if the two clump catalogs contain similar objects with similar 
masses (or fluxes).

	In the more distant future, the Cornell Caltech Atacama 
Telescope (CCAT, \citealt{ccat}) promises to be a powerful tool for 
measuring the clump mass function.  As a single-dish telescope, it will 
not suffer the spatial filtering effects that may make its contemporary, 
the Atacama Large Millimeter Array, less useful for measuring the clump 
mass function.  With an expected dish diameter of 25~m, CCAT's angular 
resolution of 2\arcsec\ at 200~\um\ will make it a powerful 
clump-finding tool.

\begin{acknowledgements}

The work made use of facilities of the Shared Hierarchical Academic
Research Computing Network (SHARCNET:www.sharcnet.ca) and Compute/Calcul
Canada.

\end{acknowledgements}

\end{document}